\documentclass[twocolumn,secnumroman,amssymb, amsmath, nofootinbib,tightenlines,
nobibnotes, aps, prb, showpacs]{revtex4}

\usepackage{graphicx}%
\usepackage{dcolumn}
\usepackage{amsmath}

\begin{document}

\preprint{HEP/123-qed}

\title{Magnetic and Electronic Properties of Li$_x$CoO$_2$ Single Crystals
}

\author{K. Miyoshi,$^1$ C. Iwai,$^1$ H. Kondo,$^1$ M. Miura,$^1$ S. Nishigori,$^2$ and J. Takeuchi$^1$}
\affiliation{%
$^1$Department of Material Science, Shimane University, Matsue 690-8504, Japan
}%
\affiliation{%
$^2$ Department of Materials Analysis, CIRS, Shimane University, Matsue 690-8504, Japan
}%
\date{\today}
\begin{abstract}
Measurements of electrical resistivity ($\rho$), DC magnetization ($M$) and specific heat ($C$) have been 
performed on layered oxide Li$_x$CoO$_2$ (0.25$\leq$$x$$\leq$0.99) using single crystal specimens. 
The $\rho$ versus temperature ($T$) curve for $x$=0.90 and 0.99 is found to be insulating 
but a metallic behavior is observed for 0.25$\leq$$x$$\leq$0.71. 
At $T_{\rm S}$$\sim$155 K, a sharp anomaly is observed in the $\rho$$-$$T$, $M$$-$$T$ and $C$$/$$T$$-$$T$ 
curves for $x$=0.66 with thermal hysteresis, indicating the first-order character of the 
transition. The transition at $T_{\rm S}$$\sim$155 K is observed for the wide range of $x$=0.46$-$0.71.  
It is found that the $M$$-$$T$ curve measured after rapid cool 
becomes different from that after slow cool below $T_{\rm F}$, which is $\sim$130 K for $x$=0.46$-$0.71. 
$T_{\rm F}$ is found to agree with 
the temperature at which the motional narrowing in the $^7$Li NMR line width is 
observed, indicating that the Li ions stop diffusing and order at the regular site below $T_{\rm F}$. 
The ordering of Li ions below $T_{\rm F}$$\sim$130 K is likely to be triggered and stabilized by the 
charge ordering in CoO$_2$ layers below $T_{\rm S}$. 

\end{abstract}

\maketitle

\section{Introduction}
\label{sec:level1}
Layered oxide Li$_x$CoO$_2$ has been intensively studied 
for the practical use as a high energy density cathode material in commercial Li ion batteries. 
Li$_x$CoO$_2$ consists of CoO$_2$ layers 
and interlayers of Li atoms alternatively stacked along $c$ axis. 
The related material Na$_x$CoO$_2$, which is originally known as a large thermoelectric material,\cite{terasaki} 
has been a subject of intensive study since the discovery of superconductivity in 
Na$_{\rm x}$CoO$_2$$\cdot$1.3H$_2$O below $T_{\rm c}$$\sim$5 K.\cite{takeda} 
In CoO$_2$ layers, Co ions are in a mixed valence state nominally having spins 
of $S$=0 (Co$^{3+}$) and $S$=1$/$2 (Co$^{4+}$), 
and form a two-dimensional (2D) regular triangular lattice, where novel types of electronic behavior are expected 
due to the geometrical frustration. 
Indeed, Na$_x$CoO$_2$ exhibits a rich variety of intriguing electronic properties as Na content $x$ increases, 
that is, superconductivity in the hydrated compound ($x$$\sim$0.35),\cite{takeda} 
an insulating ground state induced by charge ordering ($x$=0.5),\cite{foo,yokoi,gasparovic,ning} 
unusual large thermoelectric power with metallic conductivity ($x$$\sim$0.7),\cite{terasaki} 
a mass-enhanced Fermi liquid ground state analogous to LiV$_2$O$_4$ ($x$$\sim$0.75),\cite{miyoshi} 
and a spin-density-wave state ($x$$\geq$0.75).\cite{motohashi} 

Li$_x$CoO$_2$ has a similar structure with different stacking sequence of 
oxygen atom layers, where Li ions occupy the octahedral sites with three CoO$_2$ sheets per unit cell, 
while Na ions in Na$_x$CoO$_2$ occupy the prismatic sites with two CoO$_2$ sheets per unit cell. 
Li deintercalation from LiCoO$_2$ generates Co$^{4+}$ ($t^{5}_{\rm 2g}$) ions 
in the 2D triangular lattice of Co$^{3+}$ ($t^{6}_{\rm 2g}$) and leads to a hole doping 
in the $t_{\rm 2g}$ orbitals. Although LiCoO$_2$ is known to be insulating, 
a metallic behavior is therefore expected for the delithiated compound Li$_x$CoO$_2$. 
It has been reported that electrical resistivity near room temperature decreases abruptly
with decreasing $x$ from 1 to $\sim$0.9,\cite{menetrier,kellerman} 
indicating a transition from insulator to metal. 
For Li$_x$CoO$_2$, a sharp decrease in dc magnetization\cite{kellerman,sugiyama,mukai,hertz,miyoshi2,motohashi2} 
and increase in resistivity\cite{miyoshi} have been observed below 160$-$170 K 
with wide range of Li content. 
This transition is of first-order\cite{miyoshi2,motohashi2} and the occurrence of 
charge ordering in CoO$_2$ layers has been suggested.\cite{mukai,motohashi2} 
The fully delithiated compound CoO$_2$ ($x$=0), which is also the end member of Na$_x$CoO$_2$,  
has been found to be a Pauli paramagnetic metal without electron correlation due to 
the lack of two dimensionality in the structure contrasting to A$_x$CoO$_2$ (A=Li, Na).\cite{kawasaki} 
Some attempts to establish the electronic phase diagram of Li$_x$CoO$_2$ have been made 
through the synthesis for the full or nearly full range of $x$ 
using an electrochemical reaction technique.\cite{mukai,motohashi2} It is suggested that 
the critical Li content $x_{\rm c}$ at which the magnetic property changes from Pauli paramagnetic type 
to Curie-Weiss one is 0.35$-$0.45,\cite{motohashi2} whereas 
$x_{\rm c}$ for Na$_x$CoO$_2$ is known to be 0.5.\cite{foo} 

In Li$_x$CoO$_2$, various unconventional nature may be discovered as in Na$_x$CoO$_2$ due to 
the characteristic triangular CoO$_2$ layers included in common in A$_x$CoO$_2$. 
Thus, it is interesting to explore the electronic behaviors of Li$_x$CoO$_2$, 
and it also allows to throw further light on the origin of the intriguing properties of Na$_x$CoO$_2$. 
Despite the recent intensive studies, details of the intrinsic properties of Li$_x$CoO$_2$ still remain unclear. 
The investigations on Li$_x$CoO$_2$ have been so far limited to the powder samples.
For the further understanding, detailed investigations on the microscopic nature using single-crystal specimens are 
highly desirable in the future study. For the purpose, it is an important first step  
to synthesize single-crystal specimens of Li$_x$CoO$_2$ with systematic change of $x$ and 
survey the physical properties. In the present work, we have successfully synthesized single crystals of 
Li$_x$CoO$_2$ with $x$=0.25$-$0.99 and performed the measurements of 
dc magnetization, electrical resistivity and specific heat. 
A first-order transition has been observed at $T_{\rm S}$$\sim$155 K for the specimen with $x$=0.46$-$0.71. 
We demonstrate that the transition at $T_{\rm S}$ is always followed by 
the ordering of Li ions below $T_{\rm F}$$\sim$130 K. 
This suggests the possibility that the ordering of Li ions is triggered and stabilized 
by the charge ordering in CoO$_2$ layers below $T_{\rm S}$ via Coulomb coupling. 

\section{Experiment}
\label{sec:level1}
Single crystal specimens of Li$_x$CoO$_2$ were prepared by chemically extracting lithium 
from LiCoO$_2$ single crystals reacting with an oxidizer NO$_2$BF$_4$ in an acetonitrile medium. 
In the previous work, single crystals of LiCoO$_2$ were grown in an optical floating-zone furnace,\cite{miyoshi2} 
whereas in the present work those were obtained by an ion exchange reaction between Li$_2$CO$_3$ and 
Na$_{0.75}$CoO$_2$ single crystals, which are easier to obtain than LiCoO$_2$ single crystals. 
The single-crystal growth of Na$_{0.75}$CoO$_2$ was performed in a similar manner as described in the literature.\cite{chou}
Single crystals of Na$_{0.75}$CoO$_2$ cleaved to a thickness of $\sim$0.2 mm were embedded in Li$_2$CO$_3$ powder 
in an alumina boat and heated in air at 600$^{\circ}{\rm C}$ for 24 h to exchange Na and Li ions. 
An ion exchange reaction with LiNO$_3$ (250$^{\circ}{\rm C}$, 48 h) is also useful to obtain LiCoO$_2$ single crystals. 
After the reaction, the crystals were washed with distilled water repeatedly to remove Li$_2$CO$_3$ and Na$_2$CO$_3$. 
Chemical extraction of lithium from LiCoO$_2$ single crystals was carried out by immersing 
the crystals in an acetonitrile solution of NO$_2$BF$_4$ under argon atmosphere at 25$-$35 $^{\circ}{\rm C}$ for 3 days.   
Molar ratio between LiCoO$_2$ and NO$_2$BF$_4$ was changed to control the Li content $x$ ranging up to 1$:$2. 
The final products were washed with acetonitrile to remove LiBF$_4$. 
Synthesis of powder samples of Li$_x$CoO$_2$ by delithiation using NO$_2$BF$_4$ has been reported in the 
literature.\cite{venkatraman} The Li content in the pristine and delithiated specimens was 
determined by inductively coupled plasma-atomic emission spectroscopy (ICP-AES). 
In this study, we used Li$_x$CoO$_2$ single-crystal specimens with $x$=0.99 (pristine), 0.90, 
0.71, 0.66, 0.46 and 0.25. To confirm the phase purity of the specimens, 
powder X-ray diffraction (XRD) was performed by a conventional diffractometer (RINT2200, Rigaku) using Cu K$\alpha$ radiation. 
Electrical resistivity was measured using a standard dc four-probe technique. 
dc magnetization measurements were carried out using superconducting quantum interference device 
magnetometer (MPMS, Quantum Design). Specific heat was measured by a thermal relaxation method on the physical property 
measurement system (PPMS, Quantum Design).  

\begin{figure}[h]
\includegraphics[width=8cm]{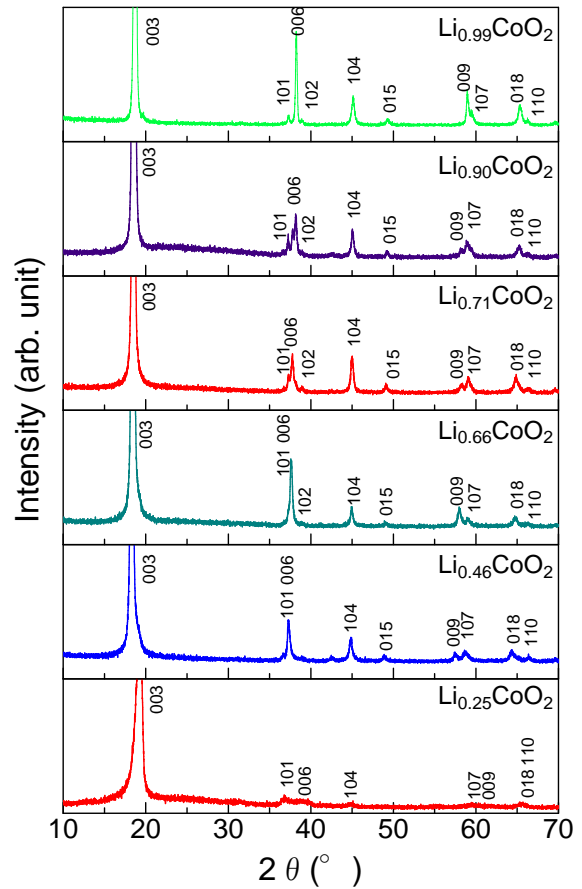}
\caption{(Color online) Powder X-ray diffraction patterns for Li$_x$CoO$_2$ with $x$=0.25$-$0.99. 
}
\label{autonum}
\end{figure}
\begin{figure}[h]
\includegraphics[width=7cm]{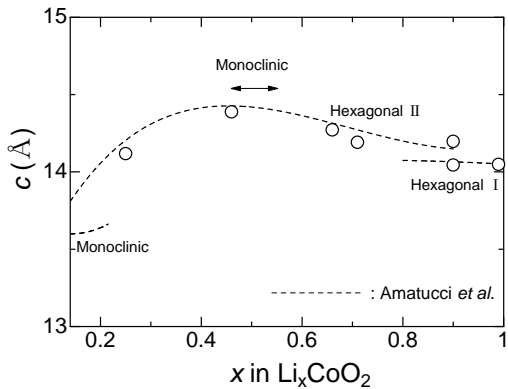}
\caption{Lattice parameter $c$ for Li$_x$CoO$_2$ with $x$=0.25$-$0.99. 
The dotted lines show 
the variation of $c$ for Li$_x$CoO$_2$ observed by in-situ X-ray diffraction 
during the electrochemical Li deintercalation by Amatucci $et$ $al$ (Ref. 21). 
Hexagonal and monoclinic phases appear in the oxidation process. 
}
\label{autonum}
\end{figure}
\begin{figure}[!]
\includegraphics[width=7cm]{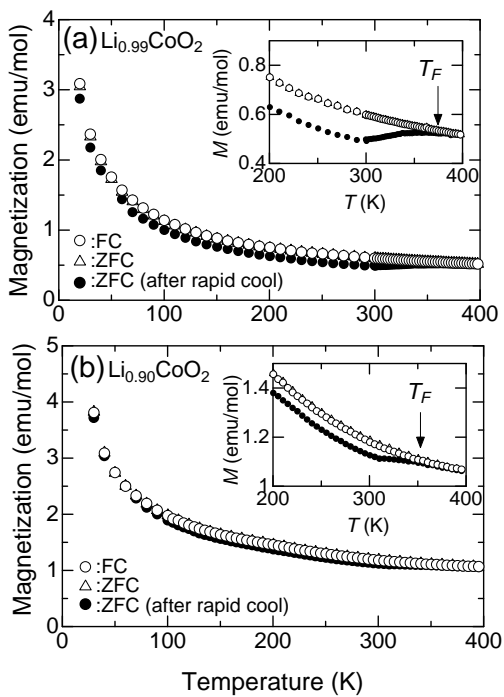}
\caption{Temperature dependence of dc magnetization 
measured with a magnetic field of $H$=1 T parallel to $ab$-plane for Li$_x$CoO$_2$ 
with $x$=0.99 (a) and 0.90 (b). 
The measurements were done after rapid cool from 
above 400 K to 10 K (closed circles) in addition to the measurements 
after slow cool down to 10 K (open symbols). 
The insets show close ups of the data for 200$\leq$$T$$\leq$400 K. 
}
\label{autonum}
\end{figure}
\begin{figure}[!]
\includegraphics[width=6cm]{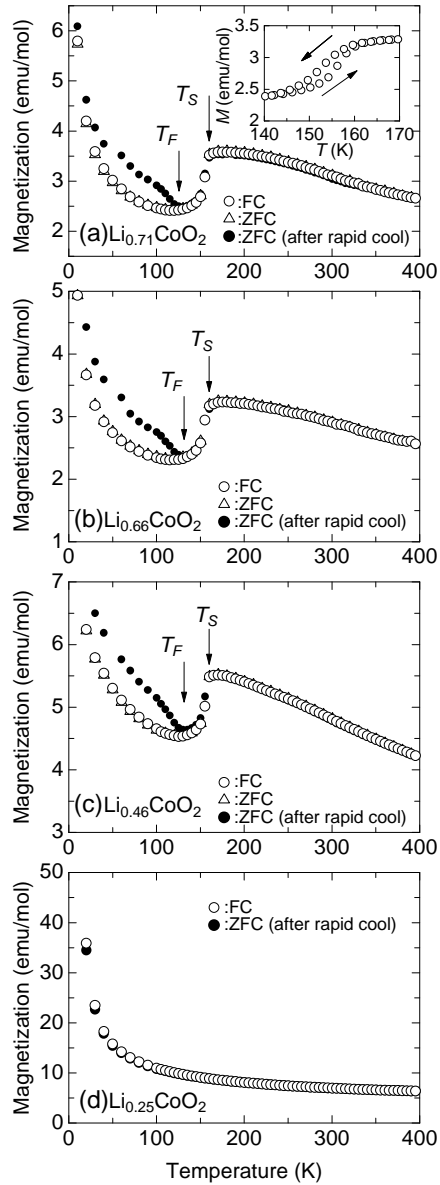}
\caption{Temperature dependence of dc magnetization measured with a magnetic field of 
$H$=1 T parallel to $ab$-plane for Li$_x$CoO$_2$  
with $x$=0.71 (a), 0.66 (b), 0.46 (c) and 0.25 (d). 
The measurements were done after rapid cool 
from 400 K to 10 K (closed circles) in addition to the measurements 
after slow cool down to 10 K (open symbols). The inset shows the data collected during 
heating and cooling between 140 K and 170 K for $x$=0.66. 
}
\label{autonum}
\end{figure}

\section{Results and Discussion}
\label{sec:level1}
\subsection{\label{sec:level2}Powder X-ray diffraction}
In Fig. 1, powder XRD patterns for Li$_x$CoO$_2$ with $x$=0.25-0.99 are shown. 
Almost all of the peaks observed in XRD are indexed based on rhombohedral type 
space group $R\bar{3}m$, indicating that the samples are of single phase, 
although a splitting of the peak into two is readily observed in 
(003), (006) and (009) reflections for the specimen with $x$=0.90.   
The peak split is due to the existence of two hexagonal phase region 
for 0.75$\leq$$x$$\leq$0.93, which has been widely observed in the electrochemical oxidation process 
by earlier workers.\cite{menetrier,motohashi2,reimers,ohzuku,amatucci} 
Figure 2 displays plots of lattice parameter $c$ versus $x$ together with 
the variation in $c$ during the electrochemical Li deintercalation 
observed by in-situ X-ray diffraction reproduced from Ref. 21. 
As seen in the figure, the variation of lattice parameter $c$ of our specimens is consistent with that 
determined in the previous study. 
As the oxidation of Li$_x$CoO$_2$ proceeds, a monoclinic phase appears in the neighborhood of $x$=0.5 
after the transition from hexagonal I to II for 0.75$\leq$$x$$\leq$0.93, 
and hexagonal and monoclinic phases coexist for $x$$\alt$0.20.\cite{reimers,ohzuku,amatucci}  
We note that more detailed structural phase diagram including the region for $x$$<$0.2 
has been recently proposed through the measurements of 
quasi-open-circuit-voltage of Li$_x$CoO$_2$$/$Al cell.\cite{motohashi2}  

\subsection{\label{sec:level2}dc magnetization}
In Figs. 3(a) and 3(b), we show the results of the dc magnetization ($M$) measurements 
as a function of temperature ($T$) for Li$_x$CoO$_2$ 
with $x$=0.99 and 0.90 measured with an applied magnetic field of $H$=1 T parallel to $ab$-plane 
under zero-field-cooled (ZFC) and field-cooled (FC) condition. 
In the measurements, dc magnetization was first measured up to 400 K after rapid cool of the sample 
from above 400 K to 10 K in zero field. For rapid cool, 
the sample was once heated above 400 K in an electric furnace and quenched in acetonitrile at 
room temperature. This was followed by rapid sample insertion 
into the chamber of the magnetometer kept at 10 K. 
Then, the measurements were performed again 
after slow cool from 400 K to 10 K at a rate $-$10 K$/$min. 
in zero field and in a magnetic field of 1 T. As seen in Figs. 3(a) and 3(b), 
the $M$($T$) curve after slow cool both for $x$=0.99 and 0.90 appears to show a Curie-Weiss type paramagnetic behavior 
and any difference is not observed in the FC and ZFC measurements, indicating the absence of 
magnetic order. On the other hand, the $M$($T$) curve after rapid cool for $x$=0.99 becomes 
different from those after slow cool below $T_{\rm F}$$\sim$380 K, as seen in the inset of 
Fig. 3(a). Also, it is found in the inset of Fig. 3(b) that the $M$($T$) curve for $x$=0.90 
becomes thermal history dependent below $T_{\rm F}$$\sim$350 K. 

In Figs. 4(a)-4(d), the $M$($T$) curves for Li$_x$CoO$_2$ with $x$=0.25$-$0.71 are shown. 
The measurements for $x$=0.25$-$0.71 were also performed both after rapid and slow cool 
in a similar manner to those for $x$=0.90 and 0.99. In Fig. 4(a), the $M$($T$) curves for $x$=0.71 exhibit 
a sharp decrease below $T_{\rm S}$$\sim$155 K. As shown in the inset of Fig. 3(a), 
a thermal hysteresis is observed in the $M$($T$) curve for $x$=0.71 around $T_{\rm S}$, indicating 
the transition at $T_{\rm S}$ is of first-order. These behaviors have 
been similarly observed in earlier studies.\cite{kellerman,sugiyama,mukai,hertz,miyoshi2,motohashi2} 
Also it is found that there is no difference in the $M_{\rm FC}$($T$) and $M_{\rm ZFC}$($T$) curves after slow cool, 
but the $M$($T$) curves after slow and rapid cool exhibit a splitting below $T_{\rm F}$$\sim$130 K. 
The characteristic behaviors of the $M$($T$) curve for $x$=0.71 are similarly observed 
for $x$=0.66 and 0.46 as seen in Figs. 4(b) and 4(c). It is noted that $T_{\rm S}$($\sim$155 K) 
and $T_{\rm F}$ ($\sim$130 K) appears to be independent of Li content $x$ for 0.46$\leq$$x$$\leq$0.71.   
As shown later, $T_{\rm F}$ agrees with the temperature at which the motional narrowing 
in the $^7$Li NMR line width is observed, indicating that the Li ions stop diffusing and 
stay at the regular site below $T_{\rm F}$. 
\begin{table*}[!]
\caption{Constant susceptibility $\chi_0$, Weiss temperature $\Theta$, Curie constant $C$, 
and effective magnetic moment $\mu_{\rm eff}$ for Li$_x$CoO$_2$ with $x$=0.99, 0.90 and 0.25. 
These parameters were extracted from the $M$($T$) data (10$\leq$$T$$\leq$400 K) using Eq. (1). 
}
\begin{ruledtabular}
\begin{tabular}{cccccc}
 $x$&$\chi_{0}$&$\Theta$ 
&$C$&$\mu_{\rm eff}$&$\mu_{\rm eff}$\\ 
&(emu/mol/Oe)&(K)&(emu/mol/K/Oe)&($\mu_{\rm B}$/Co)&($\mu_{\rm B}$/Co$^{\rm 4+}$)\\ \colrule
 0.99&2.95$\times$10$^{-5}$&$-$11&9.31$\times$10$^{-3}$&0.273&2.73\\
 0.90&7.33$\times$10$^{-5}$&$-$9.4&1.40$\times$10$^{-2}$&0.335&1.06\\
 0.25&4.81$\times$10$^{-4}$&$-$3.2&6.48$\times$10$^{-2}$&0.720&0.831\\
\end{tabular}
\end{ruledtabular}
 \label{table3}
 \end{table*}

In Fig. 4(d), the $M$($T$) curve for $x$=0.25 is found to exhibit no anomaly 
contrasting to the sample with 0.46$\leq$$x$$\leq$0.71 
but a Curie-Weiss-like $T$-dependence with no sign of magnetic order.  
$\chi$($T$) [=$M$($T$)$/$$H$] curve for $x$=0.99, 0.90 and 0.25 observed in the FC measurements was fitted with 
the following formula:   
\begin{equation}
\chi(T)=\chi_{0}+\frac{C}{T-\Theta},   
\end{equation}
where $\chi_{0}$, $C$ and $\Theta$ denote the $T$-independent susceptibility, the Curie constant and 
the Weiss temperature, respectively. Fitting parameters 
obtained from the fits of the $\chi$($T$) data (10$\leq$$T$$\leq$400 K) 
to Eq. (1) are summarized in Table 1. In the Table, effective magnetic moment $\mu_{\rm eff}$ per Co atom 
was calculated from $C$, considering that all of the Co atoms are equivalent, 
while $\mu_{\rm eff}$ per Co$^{\rm 4+}$ was calculated assuming that 
all of the Co$^{\rm 3+}$ ions are nonmagnetic ($S$=0). 
Although $\mu_{\rm eff}$ per Co$^{\rm 4+}$ is $\sim$2.7 $\mu_{\rm B}$ for $x$=0.99, 
$\mu_{\rm eff}$ yields 1.73 $\mu_{\rm B}$ corresponding to low spin Co$^{\rm 4+}$ ($S$=1/2) 
if we assume $x$$\sim$0.975. An error range $\Delta$$x$=$\pm$0.015 
of the composition determined by ICP may be possible for $x$=0.99. The value of 
$\mu_{\rm eff}$ per Co$^{\rm 4+}$ is reduced to $\sim$1.1 $\mu_{\rm B}$ for the sample with $x$=0.90, 
which is near the boundary between metallic and insulating states, as seen later. 
For $x$=0.25, $\mu_{\rm eff}$ per Co$^{\rm 4+}$ still has a large value of $\sim$0.83 $\mu_{\rm B}$, 
whereas it has been reported that Li$_x$CoO$_2$ obtained by electrochemical oxidation 
is Pauli paramagnetic for $x$$\alt$0.35.\cite{motohashi2} 
In the table, it is found that $\Theta$ has a small negative value and 
$\chi_{0}$ becomes larger with decreasing $x$ probably due to the increase of Pauli paramagnetic component. 
These features are also found in the electrochemically delithiated specimens.\cite{motohashi2} 

\subsection{\label{sec:level2}Electrical resistivity}
Next, we show the temperature dependence of electrical resistivity ($\rho$) for 
Li$_x$CoO$_2$ with $x$=0.25$-$0.99 in Figs. 5(a)-5(e). The data shown in the figures were collected during 
heating after slow cool down to $\sim$10 K ($\sim$$-$2 K$/$min.). For $x$=0.99, the $\rho$($T$) curve exhibits 
an insulating behavior in Fig. 5(a). The $\rho$($T$) curve for $x$=0.99 was confirmed not to strictly obey an 
Arrhenius law, having a slightly temperature depending activation energy of $E_{\rm a}$$\sim$0.03 eV for $T$=100$-$300 K 
and $\sim$0.02 eV for $T$$<$100 K. Also, an insulating behavior of $\rho$($T$) with an activation energy 
of 10$^{-2}$ eV order which varies slightly with temperature has been observed 
in Li$_x$CoO$_2$ with $x$=0.98$-$0.96 in an earlier study.\cite{menetrier} 
For $x$=0.90, the $\rho$($T$) curve is nearly temperature independent with resistivity $\sim$10$^0$ $\Omega$cm, 
which is $\sim$50 times smaller than that for $x$=0.99 at 300 K. 
In Fig. 5(b), the $\rho$($T$) curve for $x$=0.71 displays a metallic behavior with an amplitude of m$\Omega$cm order. 
Thus, the critical Li content at which the system changes from an insulator to metal 
is likely to be near below 0.90. The $\rho$($T$) curve for $x$=0.71 shows a sudden increase below $\sim$155 K, 
which corresponds to the first-order transition observed in the $M$($T$) curve at $T_{\rm S}$$\sim$155 K. 
Although a formation of charge ordered state in CoO$_2$ layers 
has been expected for $x$=2$/$3,\cite{motohashi2} no indicative of a transition into an insulating state is found 
in the $\rho$($T$) curve. However, we should note that the slope of the $\rho$($T$) curve for $x$=0.71 
becomes smaller below the jump at $T_{\rm S}$$\sim$155 K, suggesting that the transport mechanism 
changes below $T_{\rm S}$. The $\rho$($T$) curve for $x$=0.66 is qualitatively similar to that for $x$=0.71, 
as seen in Fig. 5(c). For $x$=0.66, the data collected 
during cooling are also shown. A slight thermal hysteresis is found in the jump-like anomaly at $T_{\rm S}$ 
as a result of the first-order character of the transition. The $\rho$($T$) curve for $x$=0.46 in Fig. 5(d) 
also shows an anomaly at $T_{\rm S}$$\sim$155 K but the anomaly is relatively small. 
In contrast, there is no anomaly in the $\rho$($T$) curve for $x$=0.25 in Fig. 5(e). 
Moreover, the residual resistivity of the $\rho$($T$) curve for $x$=0.25 is relatively 
small compared with that for higher $x$. The higher metallicity for $x$=0.25 may be 
associated with the absence of the transition at $T_{\rm S}$. 
\begin{figure}[b]
\includegraphics[width=8cm]{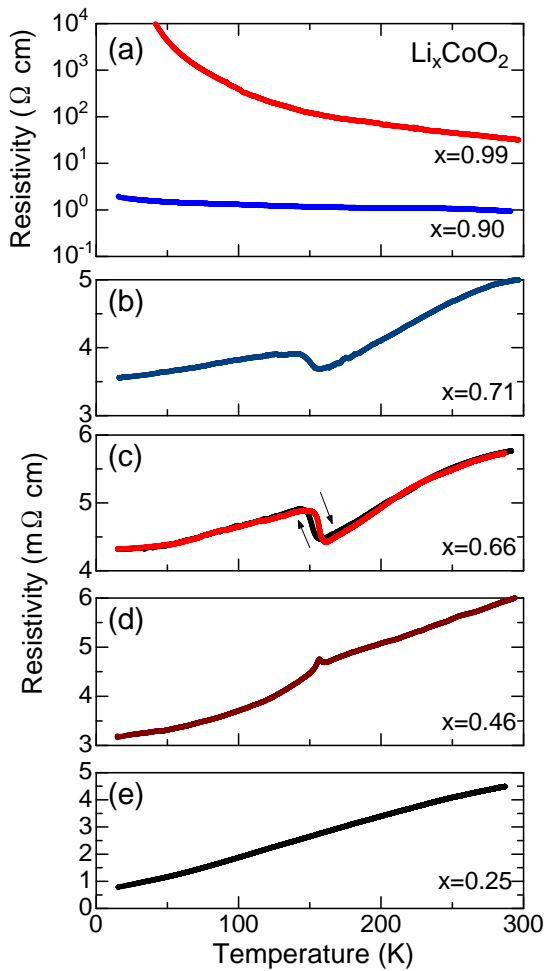}
\caption{(Color online) Temperature dependence of electrical resistivity measured along $ab$-plane 
for Li$_x$CoO$_2$ with $x$=0.99, 0.90 (a), 0.71 (b), 0.66 (c), 0.46 (d) and 0.25 (e). 
}
\label{autonum}
\end{figure}

\subsection{\label{sec:level2}Specific heat}
Figure 6 displays plots of specific heat divided by temperature $C$$/$$T$ versus 
$T$ for Li$_{0.66}$CoO$_2$. The measurements were performed during cooling and heating 
after slow cool ($-$10 K$/$min.). The data measured during heating show a 
sharp peak corresponding to the transition at $T_{\rm S}$$\sim$155 K, while 
those measured on cooling show a small peak. A large thermal hysteresis is observed 
as a strong evidence of the first-order character of the transition at $T_{\rm S}$. 
The measurements were also carried out in a magnetic field of $H$=9 T. 
The $C$/$T$ peak at $T_{\rm S}$ is found to be not field dependent, 
indicating that the transition is not magnetic one. 
Assuming the occurrence of charge ordering at the fractional Li content $x$=2/3, 
the transition entropy is calculated as $\Delta$$S$=$-$$R$(1/3 ln 1/3+2/3 ln 2/3)=5.29 J/mol K, 
which is the mixing entropy of Co$^{\rm 3+}$/Co$^{\rm 4+}$ solution.   
To examine whether or not the transition entropy at $T_{\rm S}$ for $x$=0.66 
corresponds to that of the charge ordering for $x$=2$/$3, 
we have estimated the entropy $S$ as a function of $T$ from the electronic contribution of 
specific heat $C_{\rm e}$ using the thermodynamic relationship, 
\begin{equation}
S(T)=\int_0^{T}\frac{C_{\rm e}}{T'}dT'. 
\end{equation}
$C_{\rm e}$ is given by subtracting the lattice contribution $C_{\rm lat}$ from the 
measured specific heat $C$. The $C_{\rm lat}$($T$) data were estimated from 
the $C$($T$) data for insulating Li$_{0.99}$CoO$_2$. 
In the inset of Fig. 6, the $S$($T$) curve obtained by 
integrating $C_{\rm e}$/$T$ above 2 K is shown. 
The transition entropy is estimated to be $\Delta$$S$$\sim$1.9 J/molK for $T$=140-170 K. 
This value is 36$\%$ of the transition entropy of charge ordering for $x$=2/3, 
suggesting the possibility that the charge ordering is incomplete and a part of careers are localized. 
This scenario agrees with the result that $\rho$($T$) curve for $x$=0.66 exhibits a 
sudden increase at $T_{\rm S}$ but is still metallic below $T_{\rm S}$. 
The transition entropy for Li$_x$CoO$_2$ with $x$=0.67 has been found to be $\Delta$$S$=1.49 J/molK from the magnitude of 
latent heat in the differential scanning calorimetry (DSC) measurements.\cite{motohashi2}  
In a similar context, it has been argued from the smaller $\Delta$$S$ that the charge ordering 
is incomplete and a charge disproportionation e.g., 
2Co$^{\rm +3.5}$$\rightarrow$Co$^{\rm +3.5+\delta}$+Co$^{\rm +3.5-\delta}$ occurs.\cite{motohashi2}
Also for $x$=0.71, a sharp anomaly at $T_{\rm S}$ was observed in $C$($T$)$/$$T$ data. 
For $x$=0.46, no distinct anomaly was however observed at $T_{\rm S}$. 
This leads us to remember that the anomaly in the $\rho$($T$) curve for $x$=0.46 
is relatively small compared to that for $x$=0.66 and 0.71. 
The DSC measurements have also revealed that the transition entropy for $x$=0.50 is 
$\Delta$$S$=0.47 J/molK, which is considerably smaller than that for $x$=0.67.\cite{motohashi2} 
Thus, it is inferred that the volume fraction of the charge ordering at $T_{\rm S}$ becomes continuously smaller with decreasing 
$x$ and may be negligible for $x$$<$0.40$-$0.45. 
\begin{figure}[!]
\includegraphics[width=7cm]{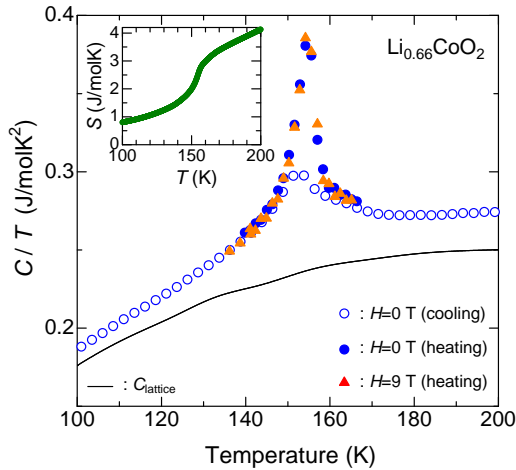}
\caption{(Color online) Plots of specific heat divided by temperature $C$$/$$T$ versus temperature $T$ 
for Li$_{0.66}$CoO$_2$ at $H$=0 and 9 T. The magnetic field was applied parallel to the $c$-axis. 
The data were collected during heating (closed symbols) 
and cooling (open circles). The solid line represents the lattice contribution. 
The inset shows magnetic entropy versus temperature curve. 
}
\label{autonum}
\end{figure}
\begin{figure}[!]
\includegraphics[width=7cm]{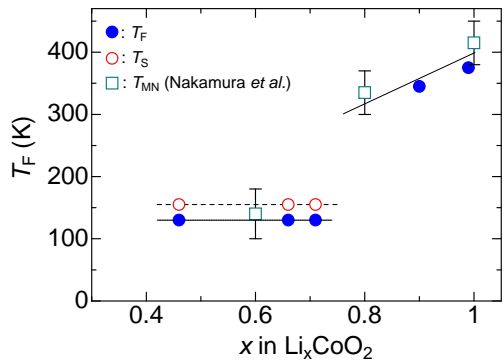}
\caption{(Color online) Plots of $T_{\rm F}$ versus $x$ for Li$_x$CoO$_2$. 
$T_{\rm S}$ and $T_{\rm MN}$ are also plotted. 
$T_{\rm MN}$ is the temperature at which the motional narrowing in the $^7$Li NMR line width has been observed (Ref. 22). 
The solid and dotted lines are guides for the eyes. 
}
\label{autonum}
\end{figure}

\subsection{\label{sec:level2}Plots of $T_{\rm F}$ versus Li content $x$}
In the dc magnetic measurements, $T_{\rm F}$, below which the $M$($T$) curve measured after 
rapid cool becomes different from that after slow cool, 
has been determined for each $x$. 
We show the plots of $T_{\rm F}$ versus $x$ in Fig. 7, where 
$T_{\rm S}$ and the temperature at which the motional narrowing in the $^7$Li NMR line width has been 
observed ($T_{\rm MN}$)\cite{nakamura} are plotted together. 
For higher $x$($\geq$0.90), $T_{\rm F}$ decreases with decreasing $x$. 
In contrast, $T_{\rm F}$ is $x$-independent for $x$$\alt$0.7.  
It is found that the characteristic relationship between $T_{\rm F}$ and $x$ 
are in good agreement with $T_{\rm MN}$ versus $x$ data, suggesting that  
$T_{\rm F}$ corresponds to the temperature above which Li ions start to diffuse from 
the regular site. This is also supported by the recent $\mu^+$SR experiments on Li$_x$CoO$_2$, 
which have revealed that the Li diffusion starts above $T^{\rm Li}_{\rm d}$$\sim$150 K both for 
$x$=0.73 and 0.53.\cite{sugiyama2} This fact is a firm evidence for 
the reliability of our result that the onset temperature of Li diffusion is $x$-independent 
for 0.45$\alt$$x$$\alt$0.70. The slight difference between $T^{\rm Li}_{\rm d}$ ($\sim$150 K) and $T_{\rm F}$ ($\sim$130 K) 
may be due to the difference of the sample. The sample used in the $\mu^+$SR experiments 
shows a sudden decrease of $M$($T$) at $\sim$170 K, which is also slightly higher than $T_{\rm S}$$\sim$155 K of our specimens. 
We have found in this study that the difference in the $M$($T$) curve after slow and rapid cool is a reliable marker to 
detect the onset of Li diffusion.  

\subsection{\label{sec:level2}Discussion}
In Li$_x$CoO$_2$, Li ions stop diffusing and the ordering of Li ions is stabilized below $T_{\rm F}$.  
The origin of the Li$^{+}$ ordering for $x$$\alt$0.7 and $x$$\agt$0.8 is considered to 
be different from each other, since $T_{\rm F}$ suddenly decreases and becomes constant for $x$$\alt$0.7. 
As seen in Fig. 7, the first order transition at $T_{\rm S}$ is always followed 
by the Li$^{+}$ ordering below $T_{\rm F}$$\sim$130 K, suggesting the possibility that 
the ordering of Li ions is triggered and stabilized by 
the charge ordering in CoO$_2$ layers below $T_{\rm S}$, where electrons combine with cobalt ions. 
This situation reminds us an intimate correlation between Co$^{3+}$$/$Co$^{4+}$ charge ordering and Na$^{+}$ ordering 
(zigzag chain) observed in Na$_{0.5}$CoO$_2$\cite{ning}, and also the characteristic ordering pattern of 
Na$^{+}$ vacancies in Na$_{0.75}$CoO$_2$ which provides an electrostatic landscape on the Co layers affecting 
the electronic and magnetic properties.\cite{roger,chou2,morris} 
For $x$=0.66, we have observed a fairly small transition entropy at $T_{\rm S}$ compared with that 
expected in the transition into charge ordered state for $x$=2$/$3, 
in addition to a metallic behavior in the $\rho$($T$) curve even below $T_{\rm S}$. 
This suggests that the carrier localization below $T_{\rm S}$ is partial. 
Indeed, an electronic disproportionation at Co sites has been observed in Na$_{0.75}$CoO$_2$, 
where only 30$\%$ of the Co site form an ordered pattern of localized Co$^{\rm 3+}$ states.\cite{julien}
Moreover, we should note that this electronic texture in the CoO$_2$ layers would contribute 
to stabilizing the ordering of Na ions,\cite{julien} as recent calculation also suggest.\cite{meng} 
Thus, the charge ordering coupled with the Li$^{+}$ ordering in Li$_x$CoO$_2$ 
may be analogous to that observed in Na$_{0.75}$CoO$_2$. 
The difference in the aspects of charge ordering between Li$_x$CoO$_2$ and Na$_x$CoO$_2$ is however obvious, since 
an insulating state is established by the complete charge ordering in Na$_{0.5}$CoO$_2$. 
To clarify the origin of the difference, 
the investigations on the ordering patterns of Li ions in Li$_x$CoO$_2$ at low $T$ and comparison with 
that in Na$_x$CoO$_2$ are highly desirable. 

Next, we focus on the behavior of $M$($T$) curve, which depends on thermal history showing a different 
amplitude of $M$ for the measurements after rapid and slow cool, as shown in Figs. 3 and 4. 
The ordering pattern of Li ions after rapid cool is likely to be different from that after slow cool  
due to the glass-like freezing of the Li$^{+}$ motions by rapid cooling, and should be 
linked with the ordering of Co$^{3+}$$/$Co$^{4+}$ via Coulomb interactions, 
which determines the amplitude of $M$. Thus, rapid cooling introduces disorder in Li layers, which 
would disturb the charge ordering in Co layers expected for $x$=0.46$-$0.71 below $T_{\rm S}$. 
The destruction of the charge ordering may lead to the enhancement of $M$ at low $T$, since the $M$($T$) 
curve exhibits a sudden increase above $T_{\rm S}$ where the charge ordering is completely destroyed. 
In Figs. 4(a)$-$4(c), the amplitude of $M$($T$) after rapid cool is indeed enhanced compared to that after slow cool 
at low $T$ and they coincide above $T_{\rm F}$, where Li ions can diffuse so that the charge ordering in 
Co layers is reproduced. 
However, we could not observe a clear anomaly in the $\rho$($T$) curve at $T_{\rm F}$ for $x$=0.46$-$0.71  
even in the measurement after rapid cool, because the number of electrons which contributes to 
the metallic conductivity never changes across $T_{\rm F}$. 
Forming a contrast, $M$($T$) measured after rapid cool is smaller than 
that after slow cool for $x$=0.90 and 0.99, as seen in Fig 3. 
In this case, the arrangement of Li ions at high $T$ also would be preserved 
by the freezing of Li$^{+}$ motions after rapid cool, 
and the difference of $M$($T$) below $T_{\rm F}$ should also originate from 
the difference in the arrangements of Li ions below $T_{\rm F}$. 
We note that the hysteresis observed for x=0.99 is rather larger than that for x=0.90. 
The larger hysteresis seems to be associated with the larger Li content in Li$_{0.99}$CoO$_2$, 
although the detailed mechanism is unclear. 
We could not rule out the possibility that Li layers are formed in an amorphous state by rapid cooling. 
If it is the case, the difference in the structure in Li layers could arise 
even for $x$$\sim$1.0 and is necessary to explain the large thermal history dependence in the $M$($T$) curve for $x$=0.99. 


\section{Summary}
\label{sec:level1}
In the present work, we have successfully obtained single crystals of Li$_x$CoO$_2$ with $x$=0.25$-$0.99 
and performed the measurements of DC magnetization, electrical resistivity and specific heat. 
It is found that the system 
changes from an insulator to metal at the critical Li content near below $x$=0.90 
with decreasing $x$. A first-order phase transition is observed 
at $T_{\rm S}$$\sim$155 K for $x$=0.46$-$0.71. 
$T_{\rm F}$ below which the $M$($T$) curve becomes thermal history dependent is found to 
correspond to the onset temperature of Li$^{+}$ diffusion. 
For $x$=0.46$-$0.71, the first-order transition at $T_{\rm S}$$\sim$155 K is always followed by  
the ordering of Li ions below $T_{\rm F}$$\sim$130 K, suggesting the possibility that 
the ordering of Li ions is triggered and stabilized by the charge ordering in Co layers. 
The transition entropy for $x$=0.66 estimated from the specific heat anomaly is found to be 1.9 mJ/molK, which is 
$\sim$36$\%$ of that for the complete Co$^{3+}$$/$Co$^{4+}$ charge ordering in Li$_{2/3}$CoO$_2$. 
In addition, the $\rho$($T$) curve for $x$=0.66 exhibits a sudden increase but still metallic below $T_{\rm S}$. 
These facts suggest that the carrier localization below $T_{\rm S}$ is in part.   
It is inferred that the volume fraction of the charge ordering below $T_{\rm S}$ becomes continuously smaller with decreasing 
$x$ and may be negligible for $x$$<$0.40$-$0.45. 
For the further understanding, microscopic investigations on Li$_x$CoO$_2$ (0.4$\alt$$x$$\alt$0.7) 
using single crystal specimens are urgently required. 
Also, it is important to establish the ordering patterns of Li ions at low $T$, 
since there would be a strong correlation between the ordering of Li$^{+}$ and the charge ordering in Co layers 
due to the electrostatic landscape provided by Li$^{+}$ ordering as in Na$_x$CoO$_2$.\cite{roger,morris} 

\begin{acknowledgments}
This work is financially supported by a Grant-in-Aid for 
Scientific Research (No. 20540355) from the Japanese Ministry of Education, 
Culture, Sports, Science and Technology. 
\end{acknowledgments}

\end{document}